\documentstyle[11pt,aaspp4]{article}

\def\sky{B_I}
\def\filtsky{B_J}
\def\skysig{\sigma_I}
\def\skysigl{\sigma_{I,\ell}}
\def\filtsig{\sigma_{J}}

\def\psfsig{\xi}
\def\psf2{\xi^2}
\def\psffunc{A}
\def\filterfunc{F}
\def\star{S_0}
\def\siq{S_I(q)}
\def\sjq{S_J(q)}
\def\crampi{C_I}
\def\crampj{C_J}
\def\gain{g}
\def\cutoff{k}
\def\rad2{( \Delta x^2  + \Delta y^2)}
\def\r2q{( [\Delta x - q]^2  + \Delta y^2)}
\def\thresh{t}
\def\p0{p_{sky}}
\def\ps{p_{obj}}
\def\betalim{\beta_{\hbox{lim}}}

\def\G3LHS{{\cal L}}

\begin{document}
\title{Cosmic Ray Rejection by Linear Filtering of Single Images\footnote{
Accepted for publication in the May 2000 issue of the {\it Publications of 
the Astronomical Society of the Pacific.}
}}

\author{James E. Rhoads}
\affil{Kitt Peak National Observatory, 950 North Cherry Avenue,
Tucson, AZ 85719\altaffilmark{2}; jrhoads@noao.edu}
\altaffiltext{2}{Current address: Space Telescope Science Institute,
3700 San Martin Drive, Baltimore, MD 21218; rhoads@stsci.edu}

\begin{abstract}
We present a convolution-based algorithm for finding cosmic rays in
single well-sampled astronomical images.  The spatial filter used is
the point spread function (approximated by a Gaussian) minus a scaled
delta function, and cosmic rays are identified by thresholding the
filtered image.  This filter searches for features with significant
power at spatial frequencies too high for legitimate objects.
Noise properties of the filtered
image are readily calculated, which allows us to compute the probability
of rejecting a pixel not contaminated by a cosmic ray (the false alarm
probability).  We demonstrate that the false alarm probability for a
pixel containing object flux will never exceed the corresponding
probability for a blank sky pixel, provided we choose the convolution
kernel appropriately.  This allows confident rejection of cosmic rays
superposed on real objects.  Identification of multiple-pixel cosmic
ray hits can be enhanced by running the algorithm iteratively,
replacing flagged pixels with the background level at each iteration.
\end{abstract}


\section{Introduction}\label{intro}
Images from most current-day astronomical instruments have tractable
noise properties.  An exemplary case is optical images from CCD
detectors, whose uncertainties are generally dominated by the Poisson
statistics of the detected photons, with (usually smaller)
contributions from detector read noise, dark current, and other
comparatively minor nuisances.  Most of these noise sources are well
approximated by Gaussian distributions, and their sum is therefore
also well approximated by a Gaussian.

Cosmic rays impinging on a detector can yield large signals over
single pixels or small groups of pixels, thereby introducing a
distinctly non-Gaussian tail to the noise distribution.  The most
common approach to removing cosmic rays from astronomical images is to
take multiple exposures and combine them with some sort of outlier
rejection.  Real astronomical objects should (usually) be present on
multiple frames, while cosmic ray hits will not generally repeat.
Such methods have been presented in the literature by (e.g.) Shaw \&
Horne (1992) and Windhorst, Franklin, \& Neuschaefer (1994), and are
widely implemented in astronomical image processing packages.

However, there are times when multiple images are not available, or
when the sources of interest may be moving or varying on timescales
short compared to the interval between exposures.  In these cases, a
cosmic ray rejection method capable of operating on single exposures
is necessary.  Cosmic ray rejection in single frames can also be
useful even when multiple exposures are to be stacked, since stacking
often requires spatial interpolation of the input images, and any
cosmic rays not previously identified can be spread over many pixels
by spatially extended interpolation kernels.  Additionally, if a stack
of images has widely different point spread function (PSF) widths,
rejection algorithms used while stacking tend either to be overly
lenient, potentially admitting cosmic rays; or overly strict,
discarding valid data from images with very good or very bad seeing.
Examples of both these behaviors are offered by sigma clipping
algorithms, where the contribution of a particular exposure to a stack
is discarded if it differs from the mean (or median) intensity at that
location by more than $k \sigma$, where $k$ is a constant (generally
with $2 \la k \la 5$) and $\sigma$ measures the intensity uncertainty
at that location.  If $\sigma$ is measured directly from the list of
exposure intensities at a fixed sky position, a lenient rejection results,
while if $\sigma$ is taken from the known Poisson statistics of electrons in
single exposures, a strict rejection results.

To identify cosmic rays in single exposures, rejection algorithms rely
on the sharpness of cosmic rays relative to true astronomical objects.
That is, any legitimate object in our astronomical image is blurred by
the PSF, but there is no such requirement on cosmic ray hits.
Provided the image is well-sampled (in practice, $\ga 2$ pixels across
the PSF full width at half maximum), cosmic ray hits can be identified
as those features with spatial variations too rapid for consistency
with the PSF.  Murtagh (1992) and Salzberg et al (1995) have explored
trainable classifier approaches to single-image cosmic ray rejection.
Their methods have the advantage of applicability to substantially
undersampled data (from the WF/PC-I instrument on the Hubble Space
Telescope).  On the other hand, these methods ultimately rely on a
training set, which may be subjectively defined.

The present paper explores a method suggested by Fischer and Kochanski
(1994), who remark that the optimal filter for detecting
[single-pixel] cosmic ray hits is the point spread function minus a
delta function.  This can be regarded as a difference between the
matched filter for detecting point sources (i.e. the PSF) and that for
detecting single pixels (i.e. a delta function).  There is one free
parameter in such a filter, which is the amplitude ratio of the two
functions.  We develop this filtering method in detail by considering
the cosmic ray rejection rates and false alarm rates.  Much of our
analysis is devoted to choosing the delta function amplitude
appropriately.  With a careful choice of this parameter, it is
possible to ensure that the false alarm rate nowhere exceeds its value
in blank sky regions.

In section~\ref{math}, we derive the noise properties of our
filtered image, and explain how to tune the filter to avoid excessive
rejection of valid data.  In section~\ref{iraf}, we
discuss practical issues that arise when implementing our algorithm.
Section~\ref{simulations} presents simulations used to verify the
algorithm's performance.
Finally, in section~\ref{theend} we summarize our work,
describe our usual application for our algorithm, and comment on a
desirable future direction for cosmic ray rejection algorithms.

\section{Mathematical formalism}\label{math}
Suppose we have an image $I$ with the following properties: First, it
has some background level $\sky$ and noise $\skysig$, and the sky
noise is uncorrelated between any pair of pixels.  Second, it is
linear in the input signal with a gain $\gain$ photons per count, so
that a pixel containing object flux $S$ counts will have a noise
contribution of $\sqrt{S/g}$ counts from Poisson noise in the object
signal.  Third, it has a point spread function that can be well
approximated by a Gaussian of characteristic width $\psfsig$ (i.e.,
the stellar profiles have a functional form $\propto \exp[-\rad2 / (2
\psf2)]$), and is well sampled (i.e., $\psfsig \ga 1$ pixel).  This
third property is an analytical convenience that is reasonably near
truth for seeing-limited optical images from ground-based telescopes.
Other PSF models would complicate the mathematical analysis that
follows, but would not greatly change either its flavor or its
quantitative results.

Now consider convolving this image with a spatial filter $\filterfunc$ 
consisting of a unit-normalized point spread function $\psffunc =
\exp[ -\rad2 / (2 \psf2) ] / (2 \pi \psf2)$ minus a scaled delta function:
$\filterfunc = \psffunc - \alpha \delta(\Delta x) \delta(\Delta y)$.
Call the convolved image $J$, so that $J=I * \filterfunc = I * \psffunc
- \alpha I$.  (Here and throughout the paper, ``$*$'' is the
convolution operator.)

If we regard the convolution kernel as a matched filter, it is clear that
a broader kernel (likely using a functional form besides the Gaussian)
would be more effective at separating cosmic rays from faint galaxies
or other extended sources.  However, almost all astronomical images
contain some legitimate pointlike sources, which should not be
rejected.  Using a template more extended than a point source would
risk rejecting stars, and such templates are therefore not explored
further.

\subsection{Noise properties of the filtered image}
We calculate the noise in the convolved image in two steps, first
determining the noise in $I * \psffunc$ and then modifying the result
to account for the second term in filter $\filterfunc$.
Treating the noise in each pixel as an independent random variable with
variance $\skysig^2$, the variance in the convolved image is simply a
weighted sum  $\filtsig^2 = \sum_\ell w_\ell^2 \skysigl^2$, where
the sum runs over pixels and $w_\ell$ is simply $\filterfunc$ evaluated
at the location $(\Delta x_\ell, \Delta y_\ell)$ of pixel $\ell$.
Now, in regions of blank sky, $\skysigl \equiv \skysig$ is constant,
so $\sigma_{I * \filterfunc} = \skysig^2 \sum_\ell w_\ell^2$.

We can calculate the noise level in $I * \psffunc$ by defining weights
$v_\ell$ as $\psffunc$ evaluated at $(\Delta x_\ell, \Delta y_\ell)$,
and noting that 
\begin{equation}
{\sigma_{I * \psffunc}^2 \over \skysig^2} =
\sum_\ell v_\ell^2 \approx \int_0^\infty { 2 \pi r dr \over (2 \pi
\psf2)^2 } \left[ \exp\left(-r^2 \over 2 \psf2 \right) \right]^2
= { 1 \over 4 \pi \psf2 } ~~,
\label{ssigAI}
\end{equation}
so that $\sigma_{I * \psffunc}^2 = \skysig^2 / ( 4 \pi \psf2)$.
The continuous approximation to the discrete sum made here should be
reasonably accurate for well-sampled data.

Modifying this for the central pixel, which has weight $w = 1/(2 \pi \psf2)
- \alpha$ rather than $v = 1/(2 \pi \psf2)$ as used above, we find
\begin{equation}
{\filtsig^2 \over \skysig^2} = \sum_\ell w_\ell^2 \approx 
 { 1 \over 4 \pi \psf2 } - \left( 1 \over 2 \pi \psf2 \right)^2 + 
 \left({ 1 \over 2 \pi \psf2 } - \alpha \right)^2 
= { 1 \over 4 \pi \psf2 } - { \alpha \over \pi \psf2 } + \alpha^2 ~~.
\label{ssigJ}
\end{equation}

\begin{figure}
\plotone{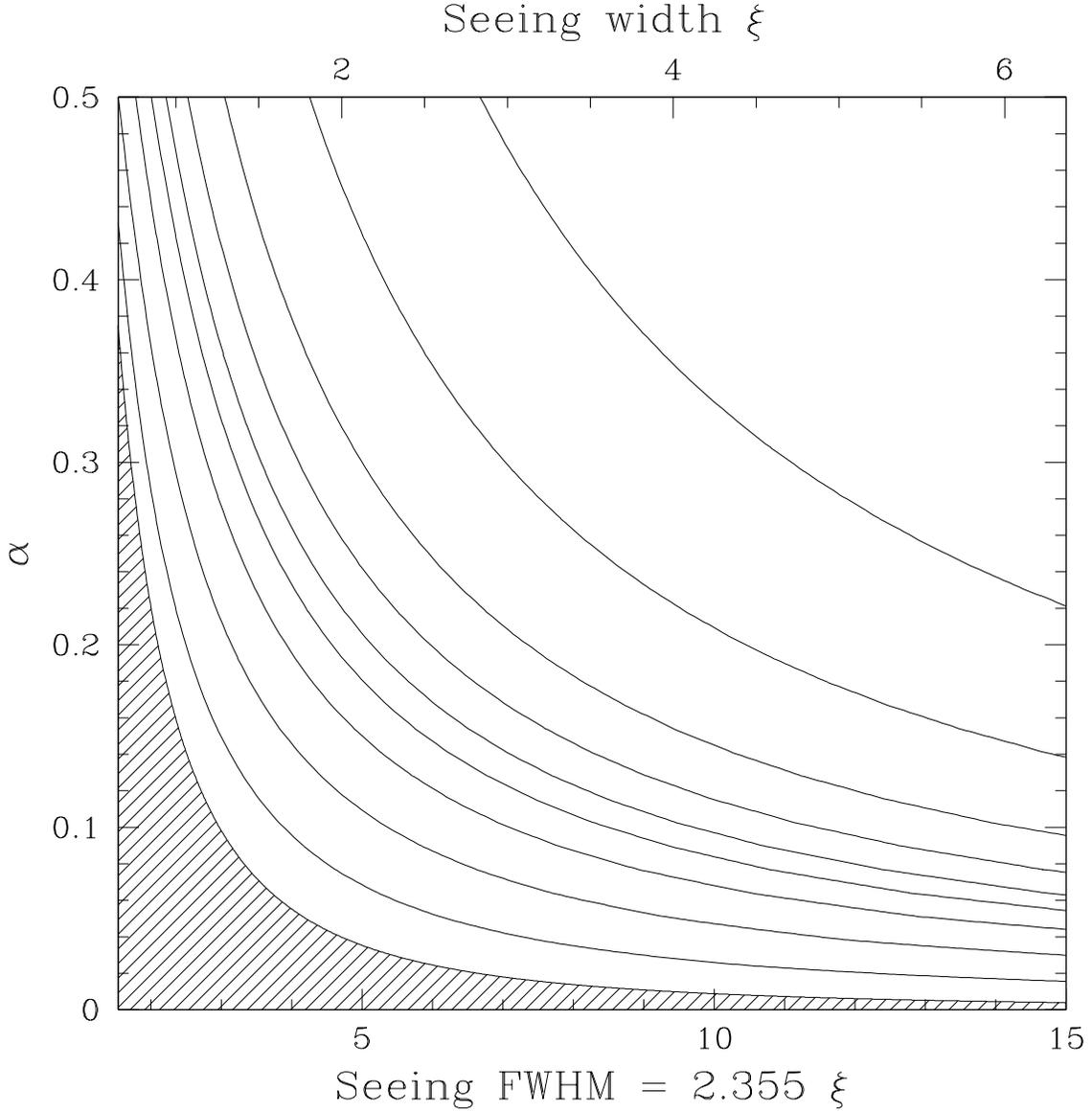}
\caption{ Contours show the factor ${ -\crampj
\skysig} \big/ {(\crampi \filtsig)}$ by which a single-pixel
cosmic ray's significance is reduced in the convolved image $J$, as a
function of the seeing and the scale factor $\alpha$ for the delta
function in the convolution kernel.
Contour levels are $0.98, 0.95, 0.90, 0.85, 0.80, 0.75, 0.667, 0.50,
0.25,$ and $0$, beginning with the top right corner and continuing to
the boundary of the hatched region.  The crosshatching indicates the
region of the parameter space where $\alpha < 1/(2 \pi \psf2)$, where
the method will necessarily fail.}
\label{crsig_a}
\end{figure}

That accomplished, we can determine the significance level that a cosmic
ray with amplitude $n \skysig$ will have in image $J$.  A single pixel
cosmic ray with $\crampi$ counts will result in a pixel with
expectation value $\crampj = \left[ 1/(2 \pi \psf2) - \alpha \right] \crampi$
below the sky level of $J$ (which is $\filtsky = (1-\alpha) \sky$).
If $ \crampi = n \skysig$, then the final significance level is 
\begin{equation}
{ -\crampj \over \filtsig} =
{\crampi \over \skysig} \times { 1/(2 \pi \psf2) - \alpha \over
\left( 1/(4 \pi \psf2) - \alpha / (\pi \psf2) + \alpha^2 \right)^{1/2} }
 = {\crampi \over \skysig} \times \left[1 + { \pi \psf2 - 1 \over (2
\pi \psf2 \alpha - 1)^2 } \right]^{-1/2} ~~.
\end{equation}
In general, this is a lower significance level than in the original
image.  In the limit of very well sampled data ($\psfsig
\rightarrow \infty$) this reduces to a significance level of 
$\crampi / \skysig$, recovering the input as one might expect.
The gradual approach to this limit simply reflects the dependence of
cosmic ray identification on the sampling of an image.
Figure~\ref{crsig_a} shows contours of ${ -\crampj  \skysig}
\big/ {(\crampi \filtsig)}$.

When we reject cosmic rays, we need to be careful not to reject the
cores of legitimate point sources.  In order to avoid doing so, we
calculate the noise level at a location near a point source, making
the same continuous approximation to discrete sums used in deriving
equation~\ref{ssigAI}.  The complication arising in this procedure is
that the presence of a source changes the noise properties of the
image.  A pixel containing object flux $I_\ell$ has noise level given
by $\skysigl^2 = \skysig^2 + I_\ell / \gain$.

We consider below the noise at a location $q$ pixels from the location
of a star with peak counts $\star$, and define function $\siq = \star
\exp\left[-q^2 / (2 \psf2) \right]$.  In the convolved image, this
pixel has expected flux
\begin{equation}
\sjq = \star \left( {1 \over 2} \exp\left[ -q^2 \over 4\psf2 \right] -
\alpha \exp\left[-q^2 \over 2\psf2 \right] \right) ~~.
\label{sjq}
\end{equation}
Since variances add linearly, modifying our earlier analysis for the
additional noise term is relatively straightforward.  For image $I *
\psffunc$, we find
\begin{eqnarray}
\lefteqn{ \sigma_{A * I}^2(q)  =   \sum_\ell v_\ell^2 \skysigl^2 } \\
& = & \sum_{\Delta x} \sum_{\Delta y}
 \left( { 1 \over 2 \pi \psf2 } \exp\left[-\rad2 \over 2 \psf2 \right]
\right)^2
\left( \skysig^2 + {\star \over \gain} \exp\left[-\r2q \over 2 \psf2
\right] \right)
\end{eqnarray}
where we have assumed (without loss of generality) that the offset to
the star is along the $x$-axis.  Defining $\Delta x' = \Delta x -
q/3$, and substituting our result from equation~\ref{ssigAI}, this becomes
\begin{eqnarray}
\sigma_{A * I}^2(q) & = & 
 { \skysig^2 \over 4 \pi \psf2 } + { \star \over (2 \pi \psf2)^2 \gain }
 \exp\left[ -q^2 \over 3 \psf2 \right]
 \sum_{\Delta x} \sum_{\Delta y} \exp\left[-3 (\Delta x'^2 + \Delta y^2)
 \over 2 \psf2 \right] \\
& \approx & { \skysig^2 \over 4 \pi \psf2 } + {\star \over (2\pi \psf2)^2
  \gain}  \exp\left[ -q^2 \over 3 \psf2 \right]
 \int_0^\infty 2 \pi r \exp\left[-3 r^2 \over 2 \psf2 \right] dr \\
 & = & { \skysig^2 \over 4 \pi \psf2 } +  \exp\left[ -q^2 \over 3 \psf2 \right]
{ \star /\gain \over 6 \pi \psf2 }
~~.
\end{eqnarray}

Again modifying the result to account for the delta function in the
convolution kernel, we obtain for the noise at distance $q$ from the
point source
\begin{eqnarray}
\lefteqn{  \filtsig^2(q) = \sigma_{A*I}^2 +
\left( \skysig^2 + \siq/\gain \right) \left(  \left[{1 \over 2 \pi
\psf2} - \alpha \right]^{2} 
  - \left[ 1 \over 2 \pi \psf2 \right]^{2} \right) } \\
 & = & \skysig^2 \left(
 { 1 \over 4 \pi \psf2 } - { \alpha \over \pi \psf2 } + \alpha^2 \right)
+ {\star \over \gain} \exp\left[-q^2 \over 2 \psf2 \right] \left(
 { \exp\left[+q^2/(6 \psf2)\right] \over 6 \pi \psf2 } -
 { \alpha \over \pi \psf2 } + \alpha^2 \right) ~~.
\label{noise2}
\end{eqnarray}
In the limit $q \rightarrow \infty$, this expression reproduces our
blank sky result (equation~\ref{ssigJ}), while at the peak of the
star, it simplifies somewhat as the exponential terms go to unity.

These results can easily be generalized to a superposition of point
sources; the second terms on the right hand sides of
equations~\ref{sjq} and~\ref{noise2} would simply be replaced by a sum
of such terms, each with its own value of the intensity parameter
$\star$ and distance parameter $q$.
% The generalization to extended sources would be more challenging,
% except insofar as extended sources can be modeled as sums of point
% sources (as is done in the CLEAN algorithm). 

\subsection{Keeping the valid peaks}
To identify cosmic rays in our image, we plan to threshold the
convolved image $J$, flagging all pixels with excessively negative
values in $J$.  There are two probabilities of interest here, namely
the probability that we will correctly flag a cosmic ray with
intensity $\crampi$ (the detection rate), and the probability that we
will incorrectly flag a pixel without cosmic ray flux (the false alarm
rate).  We have chosen to concentrate our efforts on controlling the
false alarm rate, and to accept the resulting detection rate.
From a hypothesis testing perspective (e.g., Kendall \& Stuart~1967,
chapter~22), this approach corresponds to making the null hypothesis
that a given pixel is uncontaminated by cosmic ray flux.  The false
alarm rate is then the probability of a type~I error.  Missed cosmic
ray events are type~II errors, and their probability can be calculated
as a function of cosmic ray intensity.  The tradeoffs between these
two errors for a variety of cosmic ray rejection algorithms are
reviewed by Murtagh \& Adorf (1991).

We can never set the probability of rejecting valid pixels to be
precisely zero so long as we have noise in our image and we reject any
pixels at all.  Instead we note that there is some finite probability
$\p0$ of rejecting an arbitrary sky pixel, and demand that the
probability $\ps$ of rejecting a pixel containing positive object flux
not exceed $\p0$.

Consider a threshold level in image $J$, $\thresh = -k \filtsig$.
The expected count level in $J$ is given by equation~\ref{sjq},
and the noise level there is given by equation \ref{noise2}.
We demand that
\begin{equation}
\sjq - k \filtsig(q) \ge -k \filtsig(\infty) \label{goal}
\end{equation}
in order to ensure that $\ps \le \p0$.
By using our previous expressions for $\sjq$ and $\filtsig(q)$, we
convert this into a constraint on $\alpha$.  An immediate (though
weak) constraint is that $0 \le \alpha < 1/2$, since $\filtsig(0) >
\filtsig(\infty)$, and the expected count rate must be positive to 
compensate for the increased noise at the star's location.

In the remainder of section~\ref{math}, we derive conditions
guaranteeing that inequality~\ref{goal} will hold for all values of
$\star$ and $q$.  Readers who are not interested in the mathematical
details may wish to skim section~\ref{alphsec}, which explains how to
choose the parameter $\alpha$, and then move on to section~\ref{iraf},
where we discuss implementation of the cosmic ray rejection algorithm.

By construction, condition~\ref{goal} is fulfilled as an equality for
$\star = 0$ and for any value of $q$.
To ensure that \ref{goal} holds for all $\star>0$, 
it is sufficient to show that 
\begin{equation}
{d \over d\star}\left[ \sjq - k \filtsig(q) \right] \ge 0 \label{goal2}
\end{equation}
for all $\star > 0$ and for arbitrary $q$.
Multiplying relation~\ref{goal2} by $\exp[+q^2 / (4 \psf2)]$, substituting
previous results for $\sjq$ and for $\filtsig(q)$, and
using $d \filtsig(q)^2 /  d\star = 2 \filtsig(q) \times d\filtsig(q)/d\star$,
we obtain
\begin{displaymath}
\G3LHS =
\left( {1\over 2} - \alpha \exp\left[-q^2 \over 4 \psf2 \right]
\right) \quad  -  \quad
{k\over 2 \gain} \exp\left[-q^2 \over 4 \psf2 \right] 
 \left\{ { \exp\left[+q^2/6 \psf2\right] \over 6 \pi \psf2 } -
 { \alpha \over \pi \psf2 } + \alpha^2 \right\}
 \Bigg/
\end{displaymath}
\begin{equation}
\left\{
\skysig^2 \left[ {1 \over 4 \pi \psf2 } - { \alpha \over \pi \psf2 }
+ \alpha^2 \right] + {\star \over \gain} \exp\left[-q^2 \over 2 \psf2 \right] 
\left[ {\exp\left[+q^2/6 \psf2\right]  \over 6 \pi \psf2 } -
 { \alpha \over \pi \psf2 } + \alpha^2 \right]
\right\}^{1/2} 
\ge 0 \label{goal3}
\end{equation}
for all $\star > 0$.

We now assert that for well-behaved images, it is possible to choose
$\alpha$ so that relation~\ref{goal3} is fulfilled as an equality for
$\star=0$ and $q=0$.  We will justify this assertion in
section~\ref{alphsec} below.

Taking as a hypothesis for now that $\G3LHS = 0$ for $\star=0$ and
$q=0$, we first examine the case $\star=0$, $q > 0$.  Substituting
$\star=0$ in equation~\ref{goal3} and rearranging,
\begin{displaymath}
1/2 \ge \exp\left[ -q^2 \over 4 \psf2 \right] \left\{ \alpha + { \cutoff \over
2 \gain \skysig } \left( \alpha^2 - {\alpha \over \pi \psf2} \right)
\Bigg/ \sqrt{ {1\over 4 \pi \psf2 } -{ \alpha \over \pi \psf2 } +
\alpha^2 } \right\}
\end{displaymath}
\begin{equation} 
+ \exp\left[ -q^2 \over 12 \psf2 \right] \cutoff \Bigg/ \left\{ 12 \pi
\gain \skysig \psf2  \sqrt{ {1\over 4 \pi \psf2 } -{ \alpha \over \pi
\psf2 }  + \alpha^2 } \right\} ~~.
\label{goal4}
\end{equation}
We see that the right hand side contains two exponentially decreasing
terms.  Provided both are positive, the right hand side of
equation~\ref{goal4} will clearly be a decreasing function of $q$ for
all $q\ge 0$.  The second term is positive since all of its factors
are positive by definition.  Now, by hypothesis, relation~\ref{goal4}
is an equality for $q=0$, so that the first term in~\ref{goal4} will also
be positive provided that the second is $< 1/2$ for $q=0$.  This
yields a quadratic constraint on $\alpha$:
\begin{equation}
\alpha^2 - { \alpha \over \pi \psf2 } + {1 \over 4 \pi \psf2} >
\left[ k \over 6 \pi \gain \skysig \psf2 \right]^2
  \label{goal5}
\end{equation}
This now becomes our sufficient condition for 
inequality~\ref{goal4} to be fulfilled for all $q$.

Turning our attention to $\star>0$, 
we observe by inspecting~\ref{goal3} that $\G3LHS$ is an increasing
function of $\star$ for any fixed $q$ provided only that
\begin{equation}
 \alpha^2  - { \alpha \over \pi \psf2 } + {1 \over 6 \pi \psf2 } > 0 ~~.
\label{goal6}
\end{equation}

Thus, if we can find a value of $\alpha$ that simultaneously fulfills
relation~\ref{goal3} as an equality, and fulfills
inequalities~\ref{goal5} and~\ref{goal6}, we have a convolution kernel
that will allow rejection of cosmic rays without any excess risk of
rejecting a valid pixel just because it contains flux from an object.
In the next section, we determine the parameter space over which this
is possible.

\subsection{The choice of $\alpha$}\label{alphsec}
We now turn to deriving the value of $\alpha$ that fulfills our
earlier assertion, satisfying \ref{goal3} as an equality for
$\star=0$ and $q=0$.  This is easier if we first define the auxiliary parameter
$\beta = 1/2 - \alpha$.  The intuitive significance of $\beta$ is that
the expected counts in the filtered image $J$ are $S_J(0) = \beta \star$ at
the location of a star having $\star$ counts in original image $I$.
Substituting $1/2-\beta$ for $\alpha$,
setting $\star$ and $q$ to zero, and requiring exact equality,
expression~\ref{goal3} becomes
\begin{equation}
\beta = {k \over 2 \gain \skysig}
 { \left(1 - {1\over \pi \psf2} \right) \left( {1\over 4} - \beta
 \right) + \beta^2 - { 1 \over 12 \pi \psf2}
 \over \sqrt{ \left(1 - {1\over \pi \psf2} \right) \left( {1\over 4}
  - \beta \right) + \beta^2 }  } ~~.
\label{betaiter} \end{equation}
This equation can be rearranged into a quartic in $\beta$ (or
equivalently $\alpha$).  Rather than doing so, we note that the
present version can be solved iteratively for $\beta$ by calculating
the right hand side a few times, inserting $\beta=0.25$ (or indeed any
number in $[0,0.25]$) the first time and using the previous result at
each successive iteration. $\beta$ is effectively a function of two
parameters, $k/(\gain \skysig)$ and $\psfsig$.  Now, for most imaging
CCD data, we expect reasonable choices of these parameters to be $3
\la k \la 5$, $1 \la \gain \la 10$, $\skysig \gg 5/\gain$, and
$\psfsig \ga 1$ pixel.  (Our estimate for $\skysig$ is based on the
assumption that the read noise is $\ga 5$ electrons and the observer
will typically ensure that sky noise is greater than read noise.)
This leads to $k/(\gain \skysig) \ll 1$ under typical circumstances.
In this limit, we expect
\begin{equation}
\beta \approx \betalim = {1\over 2} \left(k \over \gain \skysig \right) { 1/4 -
1/(3 \pi \psf2) \over 1/4 - 1/(4 \pi \psf2) }~~.
\label{betalim}
\end{equation}
One can further show that $\beta > \betalim$ under then nearly generic
conditions that $0 \le \beta < 1/4$ and $\beta < 1/2 - 1/(2 \pi
\psf2)$.  We therefore have a choice when implementing the algorithm
between assuming $\beta = \betalim$ or solving equation~\ref{betaiter}
iteratively.

\begin{figure}
\plotone{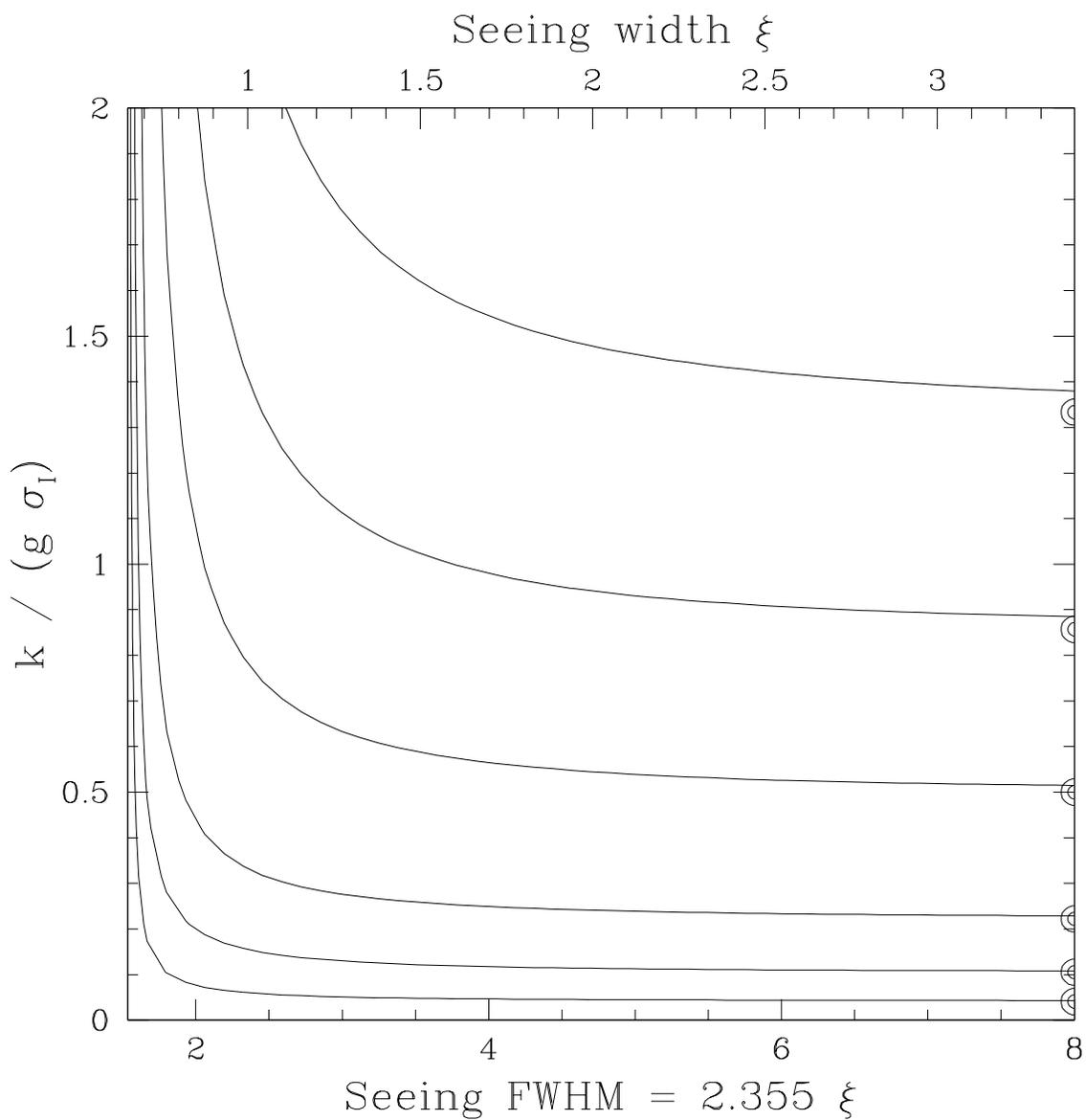}
\caption{ Contours show the optimal choice of $\beta =
1/2 - \alpha$ as a function of seeing $\psfsig$ and the rejection
threshold to noise ratio $k / (\gain \skysig)$.  Contour levels (from
top to bottom) are $0.20, 0.15, 0.10, 0.05, 0.025$, and $0.01$.
Concentric semicircles at the right hand edge show the asymptotic
value of $k / (\gain \skysig)$ for each contour curve in the limit
$\psfsig \rightarrow \infty$.}
\label{betafig}
\end{figure}

We have not derived analytically the range of parameter space over
which $\beta$ can be found iteratively, but empirically, the iterative
solution will converge to a sensible result (fulfilling
conditions~\ref{goal3}, \ref{goal5}, and~\ref{goal6}) provided that $0
< k / (\gain \skysig) < 2$ and
% that $\psfsig \ga \pi/(2 \sqrt{8 \ln(2)})$.
that $\psfsig > 2/\sqrt{3 \pi}$.
% (Of course, the continuous approximations to discrete sums that we
% made in deriving $\filtsig$ suggest that somewhat
% larger values, $\psfsig \ga 1$, are more prudent.)
These are our final set of sufficient conditions for this algorithm to
work as desired.  They are not as rigorously derived as conditions
~\ref{goal3}, \ref{goal5}, and~\ref{goal6}, but do provide a quick
check on when the method is likely to be applicable.
Figure~\ref{betafig} shows contours of $\beta$ as a function of $\psfsig$
and $k/(\gain \skysig)$.

\begin{figure}
\plotone{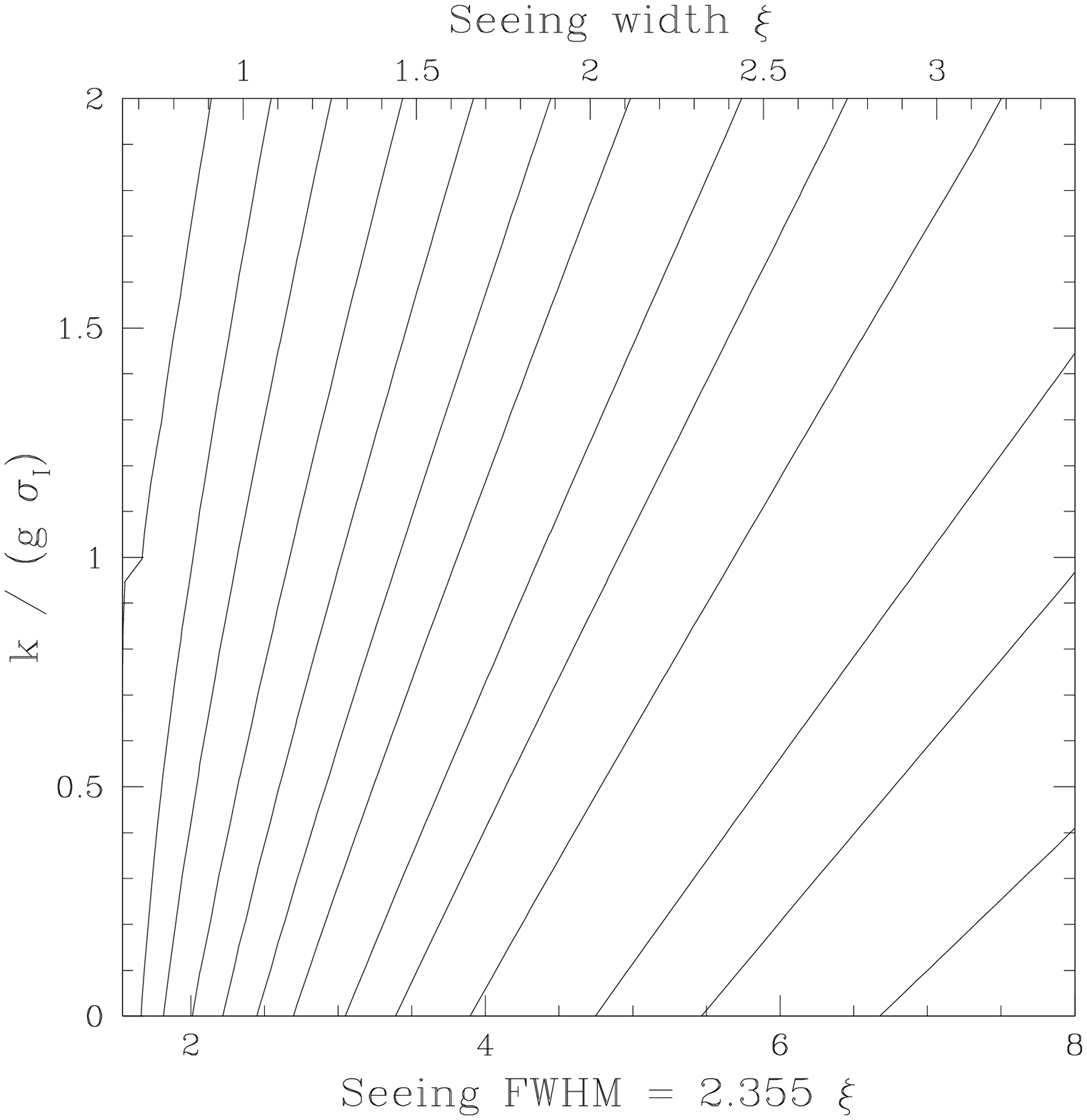}
\caption{ Contours show the factor ${ -\crampj \skysig}
\big/ {(\crampi \filtsig)}$ as a function of seeing $\psfsig$ and the
rejection threshold to noise ratio $k / (\gain \skysig)$, assuming
that the optimal choice of $\beta$ is used for the convolution kernel.
Contour levels (starting at lower right) are $0.98, 0.97, 0.96, 0.94,
0.92, 0.90, 0.87, 0.84, 0.80, 0.75, 0.68, 0.60,$ and $0.50$.}
\label{crsig_yb}
\end{figure}

Given our formula for $\beta$, it is now also possible to determine
${ -\crampj \skysig} \big/ {(\crampi \filtsig)}$
(the multiplicative reduction in significance level of a cosmic ray after
convolution) as a function of $\psfsig$ and $k/(\gain \skysig)$.
Figure~\ref{crsig_yb} shows contours of this efficiency factor.

\section{Practical Implementation} \label{iraf}
We have implemented this algorithm and applied it to several data sets
during the past year.  In doing so, we introduced several enhancements
of the basic algorithm that allow the method to run gracefully on our
real data.  Two particular artifacts were addressed by these
enhancements.  First, some bad pixels in some CCD cameras give data
values far below the sky level $\sky$.  If left alone, such pixels
will cause many of their neighbors to be flagged as cosmic rays, since
the wings of the convolution kernel will spread a strongly negative
pixel in image $I$ over many neighboring pixels in image $J$.  Second,
cosmic ray hits are often multiple-pixel events.  In this case, a
cosmic ray pixel may shield less strongly contaminated neighboring
cosmic ray pixels from identification.  An additional complication is
that the background may not be spatially uniform, which hinders
measuring the noise level in an image and defining a sensible
threshold level for cosmic ray rejection.

The problems of low pixel values and spatially variable background
levels can both be handled with preprocessing steps applied prior to
the spatial convolution.  Nonuniform sky level can be removed by
generating and subtracting a smoothed map of the background intensity.
I have chosen to use a large spatial median filter for this background
generation, but any method working on larger spatial scales than the
largest object in the frame would work.  The main caveat is that
a spatially constant rejection threshold should not be used if the
background varies enough to introduce substantial spatial variations
in the local Poisson sky noise.  Low pixels can be flagged and
replaced before the spatial convolution, using a simple threshold
operation.  This is of course best done after any variable background
is subtracted, since the sky level should be uniform for the
thresholding operation to be well behaved.  Any previously known bad
pixels can also be replaced with the background level or an
interpolation from their good neighbors at this stage.
This approach to background estimation has yielded good results for
our data, in which large scale intensity variations are weak ($< 10\%$
of $\sky$) and due primarily to flatfielding errors.  For cases where
the intensity level contains structure on a wide range of spatial
scales, multiscale transform methods (Starck, Murtagh, \& Bijaoui
1995) can provide a natural treatment of the background level.

We estimate the variance in the input image empirically, using the
iteratively clipped sample variance of the background-subtracted image
to determine $\skysig$, which is taken to be spatially uniform.
Spatial variations in the noise level could become a problem for some
images.  Such variations can be handled with a minor modification to
the algorithm, by making the rejection threshold in the filtered image
depend on the locally measured noise, provided only that the noise
level variations occur on spatial scales large compared to the
convolution kernel.  Multiscale methods can again be used for accurate
noise estimation in the presence of spatially variable backgrounds or
extended objects (Starck \& Murtagh 1998).

Some multiple-pixel cosmic ray hits will be well handled by a single
convolution and flagging step, provided that they remain smaller than
the PSF and that they are sufficiently strong (with intensities
substantially exceeding the threshold for single pixel events).
However, multiple-pixel events usually contain pixels with a range of
intensities.  When two contaminated pixels of very different intensity
lie side by side, the stronger pixel will be flagged but the weaker
one will be ``shielded'' from detection by its prominent neighbor.  To
identify such ``shielded'' cosmic ray pixels, the convolution and
flagging algorithm can be run iteratively.  After every flagging
iteration, the newly identified cosmic ray pixels are replaced with
the sky level, thereby exposing their less prominent neighbors to
scrutiny.  This iterative approach is highly successful at flagging
all parts of a multi-pixel cosmic ray hit lying above the requested
detection threshold.

When setting threshold levels for rejection,
we have chosen to use empirical measures of the sky variance in both
input and convolved image for convenience.  Comparing the results of
these empirical measures to the predicted relation given by
equation~\ref{ssigJ} gave agreement at the $5$--$7\%$ level for a test case
with $3$ pixel FWHM seeing (i.e., $\psfsig = 1.27$), with the measured
variance of the convolved image slightly exceeding the prediction.
Disagreements at this level could be due to several expected effects,
e.g., the influence of real objects on the pixel histogram (which
increases after smoothing), or the continuous approximation to
discrete sums made in deriving equation~\ref{ssigJ}.

This iterative cosmic ray rejection is of course computationally
expensive when compared to basic image reduction steps like bias
subtraction and flatfielding.  Presently, eight iterations of cosmic
ray rejection for a 2048$\times$4096 pixel image requires of order 10
minutes to run on a 295 MHz Sun Ultra-30 with 248 megabytes of main
memory.  This speed could be improved by implementing the algorithm
entirely in a compiled programming language (the present
implementation being an interpreted IRAF script).
It is nevertheless fast enough that I have routinely applied the
algorithm to large data sets (tens of $2018 \times 4096 \times 8$
pixel images from the Kitt Peak National Observatory CCD Mosaic camera).
% Comment on execution time scalings?
In principle, the computational requirements should scale as $n
\log(n)$ for $n$ pixels in the large-$n$ limit, since the convolution
can be implemented using fast Fourier transforms, while the remaining
steps should all be linear in the number of pixels.  Timing tests on
a 400 MHz Intel Pentium-II computer with 128 megabytes of memory
yielded a scaling of approximately $n^{1.4}$ for images with
$\log_2(n) = 22 \pm 1$ (i.e. roughly 2k by 2k pixels).
The difference between this scaling and the $n \log(n)$ scaling suggested
from first principles is perhaps due to the variety of different
computational demands (memory, i/o, cpu speed) which can limit the
performance of the algorithm for different image sizes.

\section{Simulations} \label{simulations}
To verify the analytical results of section~\ref{math} and study the
effectiveness of the iterated algorithm on multiple pixel events, we
carried out three types of artificial data simulations.  The first
type tested the algorithm's detection rate for single pixel events;
the second tested the false alarm rate at the locations of point
sources; and the third tested rejection of multi-pixel events.  For
both tests of detection rates, the empirically measured detection
threshold was taken as the intensity of added cosmic rays for which
50\% of the affected pixels were correctly flagged.  This is the
appropriate cutoff because a simulated cosmic ray of precisely
threshold intensity will be boosted above the cutoff by Poisson noise
half the time, and will fall below threshold the other half.

The cosmic ray detection rate test added single pixel cosmic ray hits
to a noise field and counted the number of hits correctly flagged by
the algorithm as a function of CR intensity, rejection threshold, and
PSF width.  This allows a check of the analytic results plotted in
figure~\ref{crsig_yb}.  The agreement is good, with empirically
measured detection thresholds falling between 100\% and 110\% of the
theoretical expectations throughout the tested parameter space.  In
particular, the measured detection threshold is within 3\% of the
predicted value for $\cutoff / ( \gain \skysig ) \la 0.3$, which is
the regime of greatest interest for broadband astronomical imaging.

An interesting variant on the detection test is to run it on a field
with stars or other astronomical objects.  Our tests showed an
appreciable degradation ($\sim 7\%$) of the CR detection threshold
averaged over the image in the presence of a reasonably dense star
field.  This is expected, since the detection efficiency decreases in
the wings of a point source (see section~\ref{math}).  However, there
is no good way to characterize this effect for all possible images.
If a precise measurement of the detection threshold in some particular
image is needed, it can be obtained through simulations by adding
``cosmic ray hits'' to that exact image and studying their recovery rates.

The false alarm rate test examined the probability of rejecting a
given pixel in a pure Poisson noise field with and without point
sources.  Only the central pixels of the point source locations were
considered, since the wings of stellar profiles are less likely than
their cores to be incorrectly rejected.
This test confirmed that the probability of rejecting the central
pixel of a star does not exceed the probability of rejecting an
arbitrary sky pixel under our algorithm.

Finally, the multiple pixel event tests placed artificial bad columns
onto noise fields and measured the fraction of rejected pixels.  Bad
columns were chosen as a suitably conservative limiting case of
multiple-pixel cosmic ray events, since such events usually have a
linear morphology.  The realized rejection threshold was determined as
a function of stellar FWHM, number of adjacent bad columns, and
requested sigma clipping level.  These simulations were run with a sky
noise of $41$ ADU and gain of $3$, so that they are restricted to low
values of $\cutoff / ( \gain \skysig )$.  The general result is
(unsurprisingly) that features comparable in size to the PSF cannot be
rejected, while features much smaller than the PSF on only one axis
can be rejected with relatively modest increases in the intensity
threshold for rejection.  Results of the multi-pixel event
simulations are summarized in table~\ref{colrej_tab}.

\begin{table}
\begin{tabular}{l|l|lllllll}
Sigma & $N_{\hbox{cols}}$ & \multicolumn{7}{c}{Stellar FWHM} \\
cutoff & & 2 & 2.52 & 3.175 & 4 & 5.04 & 6.35 & 8 \\
\hline
3 & 1 & 0.34 & 0.57 & 0.74 & 0.80 & 0.86 & 0.90 & 0.94 \\
  & 2 & 0    & 0    & 0    & 0.40 & 0.56 & 0.70 & 0.80 \\
  & 3 & 0    & 0    & 0    & 0    & 0.19 & 0.39 & 0.58 \\
  & 4 & 0    & 0    & 0    & 0    & 0    & 0.12 & 0.31 \\
\hline
4 & 1 & 0.22 & 0.53 & 0.67 & 0.76 & 0.82 & 0.86 & 0.90 \\
  & 2 & 0    & 0    & 0    & 0.32 & 0.49 & 0.59 & 0.70 \\
  & 3 & 0    & 0    & 0    & 0    & 0.09 & 0.32 & 0.38 \\
  & 4 & 0    & 0    & 0    & 0    & 0    & 0.07 & 0.25 \\
\hline
5 & 1 & 0.16 & 0.52 & 0.63 & 0.73 & 0.81 & 0.85 & 0.89 \\
  & 2 & 0    & 0    & 0    & 0.23 & 0.43 & 0.50 & 0.63 \\
  & 3 & 0    & 0    & 0    & 0    & 0.06 & 0.27 & 0.36 \\
  & 4 & 0    & 0    & 0    & 0    & 0    & 0.04 & 0.21 \\
\hline
\end{tabular}
\caption{The reduction in cosmic ray detection sensitivity for multiple
pixel events consisting of adjacent bad columns of uniform value.  The
ratio of theoretical single pixel CR detection threshold to
empirically determined multiple pixel threshold level is tabulated as
a function of sigma cutoff, transverse size of the defect in columns,
and point spread function width.}
\label{colrej_tab}
\end{table}

\section{Summary and Discussion} \label{theend}
We have presented a cosmic ray rejection algorithm based on a
convolution of the input image.  The advantages of the method spring
from the linear nature of the spatial filter, which allows us to
determine the noise properties of the filtered image and so to
calculate and control the probability of rejecting the central pixel
(or indeed any pixel) of a point source.  This safety mechanism
ensures that cosmic ray rejection can be applied throughout the image,
without special treatment for the locations of sources.  The
sensitivity to cosmic rays is of course reduced at the locations of
objects, because of the added Poisson noise contributed by object
photons and the resulting need to maintain a positive expectation
value in the filtered image.

We usually apply our method conservatively, considering pixels
innocent until proven guilty beyond any reasonable doubt.  This means
that given some uncertainty in the measured point spread function, we
use a convolution kernel that is slightly narrower than our best
estimate of the PSF (generally by about 10\%).  This choice depends on
the relative importance of keeping legitimate sources and rejecting
spurious ones for the scientific problem at hand.

Our original goal in developing this algorithm was to flag and
replace cosmic ray hits in individual exposures that are later
aligned and stacked.  The alignment procedure requires
interpolating the original images, and we use sinc interpolation to
preserve the spatial resolution and noise properties of the input
image.  However, sinc interpolation assumes well sampled data and
responds badly to cosmic rays, spreading their effects over many more
pixels than were originally affected and motivating us to replace them
at an early stage.  We nevertheless have a second chance to reject
cosmic rays by looking for consistency among our different exposures
when we stack them, and this second chance helps motivate our generally
conservative approach to cosmic ray flagging.

By applying the algorithm developed here followed by sigma rejection
during image stacking, we exploit two distinct properties of cosmic
rays: They are sharper than the point spread function, and they do not
repeat from exposure to exposure.  However, we are using these two
tests in sequence.  An algorithm exploiting both
pieces of information simultaneously could potentially yield more sensitive
cosmic ray rejection.  For general data sets, such an algorithm would
have to handle stacks of unregistered images with different PSFs,
making its development difficult but potentially rewarding.  An
interesting effort in this regard is Freudling's (1995) algorithm,
which identifies cosmic rays in the course of deconvolving and
coadding images with Hook \& Lucy's (1992) method.

\acknowledgements
This work was supported by a Kitt Peak Postdoctoral Fellowship and
by an STScI Institute Fellowship.  Kitt
Peak National Observatory is part of the National Optical Astronomy
Observatories, operated by the Association of Universities for
Research in Astronomy (AURA) under cooperative agreement with the
National Science Foundation.  The Space Telescope Science Institute
(STScI) is operated by AURA under NASA contract NAS 5-26555.
I thank an anonymous referee for their remarks.

\end{document}